\documentclass[conference,a4paper]{IEEEtran}

\usepackage{amsmath,graphicx}
\usepackage{url}
\usepackage[hidelinks]{hyperref}
\usepackage{epsfig}
\usepackage{subcaption}
\usepackage{amsmath}
\usepackage{amssymb}
\usepackage{color}
\usepackage{url}
\usepackage{xspace}
\usepackage{multirow}

\usepackage{graphicx} 
\usepackage{subcaption}
\usepackage{url}
\usepackage{tabularx,booktabs}
\newcolumntype{Y}{>{\centering\arraybackslash}X}
\newcolumntype{R}{>{\raggedleft\arraybackslash}X}
\usepackage{arydshln}
\usepackage{tabu}
\usepackage{ctable}
\usepackage{enumerate,multicol}
\usepackage{multirow}
\usepackage{multicol}
\usepackage{enumitem}
\usepackage{amsmath,array,graphicx}
\usepackage{gensymb}
\usepackage{float}
\usepackage{caption}
\usepackage{subcaption}
\usepackage{balance}
\usepackage{xspace}
\usepackage{xcolor}
\usepackage{enumerate,multicol}
\usepackage[ruled,vlined]{algorithm2e}
\usepackage{glossaries}

\newcommand{\etal}{\emph{et al.\xspace}}

\newcommand{\etc}{\emph{etc.\xspace}}

\newacronym{qp}{QP}{Quantization Parameter}
\newacronym{rdo}{RDO}{Rate-Distortion Optimization}
\newacronym{cu}{CU}{Coding Unit}
\newacronym{vvc}{VVC}{Versatile Video Coding}
\newacronym{hevc}{HEVC}{High Efficiency Video Coding}
\newacronym{ctu}{CTU}{Coding Tree Unit}
\newacronym{avc}{AVC}{Advanced Video Coding}
\newacronym{av1}{AV1}{AOMedia Video 1}
\newacronym{cnn}{CNN}{Convolutional Neural Network}
\newacronym{isp}{ISP}{Intra Sub Partitioning}
\newacronym{geo}{GEO}{Geometric partitioning}

\newacronym{ns}{NS}{Non-Split}
\newacronym{hbt}{HBT}{Horizontal Binary Tree}
\newacronym{vbt}{VBT}{Vertical Binary Tree}
\newacronym{htt}{HTT}{Horizontal Ternary Tree}
\newacronym{vtt}{VTT}{Vertical Ternary Tree}
\newacronym{qt}{QT}{QuadTree Split}
\newacronym{psnr}{PSNR}{Peak Signal-to-Noise Ratio}
\newacronym{vvenc}{VVenC}{Versatile Video Encoder}
\newacronym{vvdec}{VVdeC}{Versatile Video Decoder}
\newacronym{vtm}{VTM}{VVC Test Model}
\newacronym{vmaf}{VMAF}{Video Multi-method Assessment Fusion}
\newacronym{rdcost}{RDcost}{Rate-Distorsion Cost}
\newacronym{ctc}{CTC}{Common Test Condition}
\newacronym{vca}{VCA}{Video Complexity Analyzer}
\newacronym{ra}{RA}{RandomAccess}
\newacronym{fps}{FPS}{Frames Per Second}

\title{Preparing VVC for Streaming: \\A Fast Multi-Rate Encoding Approach}

\iftrue
\author{
\IEEEauthorblockN{Yiqun Liu\IEEEauthorrefmark{1}, Hadi Amirpour\IEEEauthorrefmark{2}, Mohsen Abdoli\IEEEauthorrefmark{3}, Christian Timmerer\IEEEauthorrefmark{2}, and Thomas Guionnet\IEEEauthorrefmark{1}
}

\IEEEauthorblockA
{
\IEEEauthorrefmark{1}Ateme, Rennes, France
}

\IEEEauthorblockA
{
\IEEEauthorrefmark{2}Christian Doppler Laboratory ATHENA, Alpen-Adria-Universit{\"a}t, Klagenfurt, Austria
}

\IEEEauthorblockA
{
\IEEEauthorrefmark{3}IRT b$<>$com, Cesson-S\'{e}vign\'{e}, France
}

}
\fi
\begin{document}
\sloppy
\maketitle
\begin{abstract}
The integration of advanced video codecs into the streaming pipeline is growing in response to the increasing demand for high quality video content. However, the significant computational demand for advanced codecs like \gls{vvc} poses challenges for service providers, including longer encoding time and higher encoding cost. This challenge becomes even more pronounced in streaming, as the same content needs to be encoded at multiple bitrates (also known as representations) to accommodate different network conditions. To accelerate the encoding process of multiple representations of the same content in \gls{vvc}, we employ the encoding map of a single representation, known as the reference representation, and utilize its partitioning structure to accelerate the encoding of the remaining representations, referred to as dependent representations. To ensure compatibility with parallel processing, we designate the lowest bitrate representation as the reference representation. The experimental results indicate a substantial improvement in the encoding time for the dependent representations, achieving an average reduction of 40\%, while maintaining a minimal average quality drop of only 0.43 in \gls{vmaf}. This improvement is observed when utilizing \gls{vvenc}, an open and optimized \gls{vvc} encoder implementation.

\end{abstract}

\begin{IEEEkeywords}
VVC, multi-rate, VVenC, CTU, fast partitioning, random access.
\end{IEEEkeywords}

\section{Introduction}
\label{sec:introduction}
The demand for high quality video streaming is continuously rising, making the transition to more advanced video codecs inevitable. While H.264/\gls{avc}~\cite{wiegand_overview_2003} continues to maintain its dominance as the primary video codec in the industry, reports indicate a growing trend of adopting more advanced video codecs like \gls{hevc}~\cite{sullivan_overview_2012} and \gls{av1}~\cite{chen_overview_2018}. Many companies are actively adding support for these codecs in their products. For instance, Google has incorporated \gls{hevc} support into Chrome, enabling users to experience the benefits of this advanced codec while streaming videos. In addition, Meta has introduced \gls{av1} support for reel videos, enhancing the visual quality and efficiency of their video platform. These initiatives highlight adopting and leveraging the capabilities of cutting-edge video codecs for improved user experiences~\cite{bitmovin_dev_rep22_ref}.

As the latest video codec, \gls{vvc}~\cite{bross_overview_2021} showcased a remarkable reduction in bitrate of around 50\% for the same subjective video quality compared to its predecessor standard, \gls{hevc}~\cite{baroncini2020vvc,baroncini2021dry}. Wieckowski~\etal~\cite{wieckowski_vvc_2022} highlight the feasibility of employing \gls{vvc} in a streaming application through a practical workflow. The study presents the successful integration of \gls{vvenc}, an open and optimized implementation~\cite{wieckowski_vvenc_2021} of the \gls{vvc} encoder, into Bitmovin's cloud-based encoding solution. The paper thoroughly examines the effects of \gls{vvc} on practical considerations like bitrate ladder selection, cost, and performance compared to other codecs. Furthermore, it shows the utilization of \gls{vvdec} with WebAssembly, enabling seamless real-time playback of \gls{vvc} directly within web browsers.

As we move towards more advanced video codecs to achieve higher-quality video delivery, one common trade-off is the increased computational time associated with these codecs. This aspect becomes even more critical in video streaming applications, where the encoding process involves multiple representations of the same content to accommodate varying network conditions for users. For instance, \gls{vvc} demonstrates a 5x increase in complexity compared to \gls{hevc} when encoding the same content under LowDelay conditions~\cite{pakdaman_complexity_2020}. This increase in complexity is primarily attributed to the utilization of more sophisticated tools, which inherently introduce complexity. The primary contributor to the increased complexity in \gls{hevc} and \gls{vvc} is the \gls{ctu} partitioning~\cite{cetinkaya_ctu_2021,9816276}. 

There are many approaches that have been introduced to reduce the complexity of \gls{vvc}, in particular fast \gls{ctu} partitioning algorithms for the \gls{vtm} encoder. Heuristic methods make full use of pixel-related statistics, such as gradient, variance, \etc~Fan \textit{et al.} in~\cite{fan2020fast} developed a hybrid early termination mechanism based on pixel variance and gradient. Similarly, a gradient-based early termination was proposed in~\cite{cui2020gradient}.

In streaming applications, where video content is encoded at multiple bitrates, fast multi-rate encoding methods are employed. These methods typically designate a single representation as the reference representation, encoding it first, and utilizing its information to expedite the encoding process for the remaining representations, known as dependent representations. 
Schroeder~\etal~\cite{schroeder_block_2015} proposed a multi-rate encoding approach in which the video is first encoded at a high quality representation, and its information is reused to accelerate the encoding of lower quality representations. They leveraged the fact that each \gls{ctu} achieves its highest partitioning depth in the highest quality representation. Consequently, by determining the partitioning depth of \gls{ctu}s in the highest quality representation ($d_m$), it is possible to skip the partitioning of co-located \gls{ctu}s in lower quality representations beyond $d_m$. This leads to significant time savings during the encoding process. Çetinkaya~\etal~\cite{cetinkaya_fame-ml_2020,cetinkaya_fast_2021} proposed an approach in which the lower quality representation is initially encoded, and for each \gls{ctu} with a partitioning depth of ($d_m$), depths lower than $d_m$ are skipped during the \gls{rdo} search. In this method, \glspl{cnn} is used to extract additional features, enhancing the performance of the partitioning depth decisions. Amirpour~\etal~\cite{amirpour_towards_2021} investigate the trade-off between time savings and quality drop by considering different representations as reference representation. They concluded that utilizing a middle quality representation can effectively strike a balance between these factors, optimizing the trade-off. Menon~\etal~\cite{menon_emes_2022} presented an extensive investigation of various multi-rate schemes and introduced innovative heuristics aimed at constraining the \gls{rdo} process across different representations. Building upon these heuristics, they proposed three multi-encoding schemes that leverage encoder analysis sharing across various representations. These schemes were designed to optimize: \textit{(i)} highest compression efficiency, \textit{(ii)} the optimal trade-off between compression efficiency and encoding time savings, and \textit{(iii)} maximum encoding time savings.

While the aforementioned methods have demonstrated significant performance in the context of \gls{hevc}, the exploration of fast multi-rate encoding techniques in \gls{vvc} remains relatively unexplored. In this paper, we aim to bridge this gap by applying fast multi-rate encoding to \gls{vvenc}, an optimized open-source software implementation of \gls{vvc}. By leveraging this approach, we investigate the potential benefits and performance improvements of fast multi-rate encoding in the \gls{vvc} domain. In the following section, we provide a concise overview of \gls{ctu} partitioning in \gls{vvc}. Section~\ref{sec:3} outlines our proposed fast multi-rate encoding approach specifically tailored for \gls{vvc}. The experimental results are presented in Section~\ref{sec:4}. Finally, Section~\ref{sec:5} concludes the article, summarizing the key findings and contributions.


\section{VVC partitioning in practice}
\label{sec:ctu}
\subsection{Multi-Type (MTT) partitioning}
\begin{figure}[h]
    \centering
    \includegraphics[width=1\linewidth]{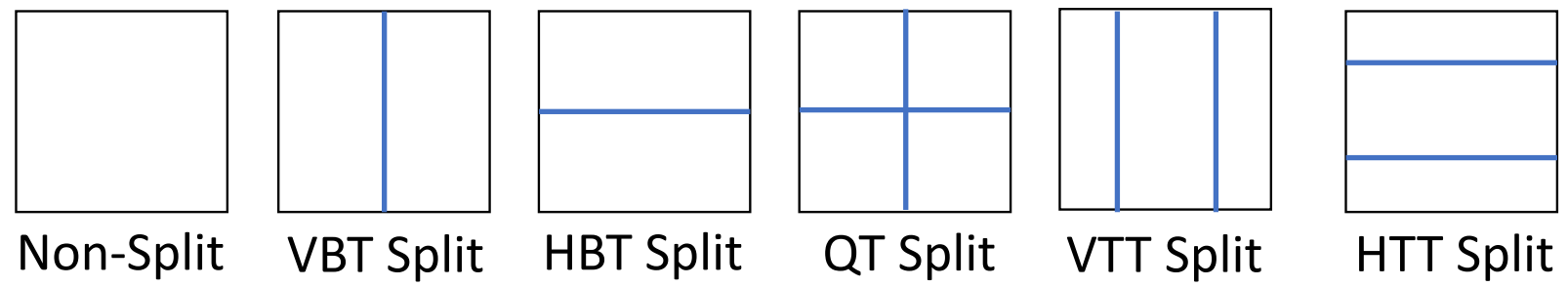}
    \caption{Different split types in VVC.}
    \label{fig:split_types}
    \vspace{-2em}
\end{figure}

\gls{vvc} offers more efficient partitioning through larger and more flexible \gls{cu} splits than \gls{hevc}~\cite{huang_block_2021}. To be precise, the largest and smallest \gls{cu}s in \gls{vvc} are 128$\times$128 and 4$\times$4 for luma samples, respectively. Six types of splits are also allowed, namely \gls{ns}, \gls{qt} split, \gls{hbt} split, \gls{vbt} split, \gls{htt} split, and \gls{vtt} split, as presented in Fig.~\ref{fig:split_types}. \gls{ns} refers to encoding the entire \gls{cu} without applying any split on it. Moreover, \gls{vvc} is integrated with tools such as \gls{isp}~\cite{isp} or \gls{geo}~\cite{geo} that are capable of further splitting \gls{cu}s into the smallest units to provide more accurate texture/motion modeling.

\subsection{The challenge of partitioning RDO search}
\vspace{-0.2em}
Partitioning is in the heart of the \gls{rdo} search; hence, it typically decides the overall encoder complexity. Precisely, the encoder is provided with such freedom of parameters, so that encoding a single video could be as fast as real-time (\textit{e.g.} 60 frames per second) or as slow as days of encoding. Depending on the underlying use-case, either one could be preferred. In particular, a slow encoding is popular in streaming services, where the video titles are typically transmitted to receivers on-demand, therefore, compression efficiency has priority to the encoding complexity. However, extremely long encoding could be costly, especially given that most encodings are performed on relatively expensive cloud servers. Therefore, even for streaming services, it is always preferred to deploy an encoder that is fast and efficient as much as possible. 

Practical encoders deploy various early termination shortcuts to reduce the complexity of \gls{rdo}. The aim of these shortcuts -- typically applied on the partitioning search -- is to avoid searches that are less likely to be optimal. Specifically, given the partitions checked in the past and the trend of their \gls{rdcost}s, the shortcuts determine whether to check each partition size. The effectiveness of this approach depends on different aspects, including texture/motion complexity, the accuracy of the \gls{rdcost} estimation, and last but not least, the encoded bitrate. In the last case, previous studies show that early termination shortcuts often skip more RD checks for lower bitrates. For instance, encoding JVET's \gls{ctc}~\cite{jvet_ctc} sequences with \gls{qp} 37 is about 80\% faster than with \gls{qp} 22 \cite{9897595}. This is mainly due to the fact that at lower bitrates, large \gls{cu}s often end up having lower \gls{rdcost}s than smaller \gls{cu}s and this trend is usually identified early enough during the top-to-bottom partitioning search to trigger and early termination that saves several partitioning checks.

A challenge in deploying effective partitioning acceleration lies in the concept of \gls{cu} depth, which is defined as the number of times a \gls{ctu} should split in order to reach a certain \gls{cu} size at a certain position in the \gls{ctu}. In \gls{hevc}, the depth of a \gls{cu} is directly determined by the size of \gls{cu}, since only quad split partitioning is available in \gls{hevc}. However, in \gls{vvc}, binary and ternary splits are designed such that the depth of a leaf rectangular \gls{cu} cannot merely be determined by its size and shape. For instance, to obtain four 8$\times$8 \glspl{cu} from a 16$\times$16 \gls{cu}, one has the freedom to perform either a single \gls{qt}, or a \gls{vbt} followed by an \gls{hbt}, or an \gls{hbt} followed by a \gls{vbt}. As a result, the use of \gls{cu} depth in \gls{vvc} to exploit partitioning size correlations is less evident than in \gls{hevc}.


\subsection{Multi-rate encoding opportunities}
\begin{figure}[!t]
    \centering
    \begin{subfigure}{0.24\textwidth}
        \centering
        \includegraphics[width=\textwidth]{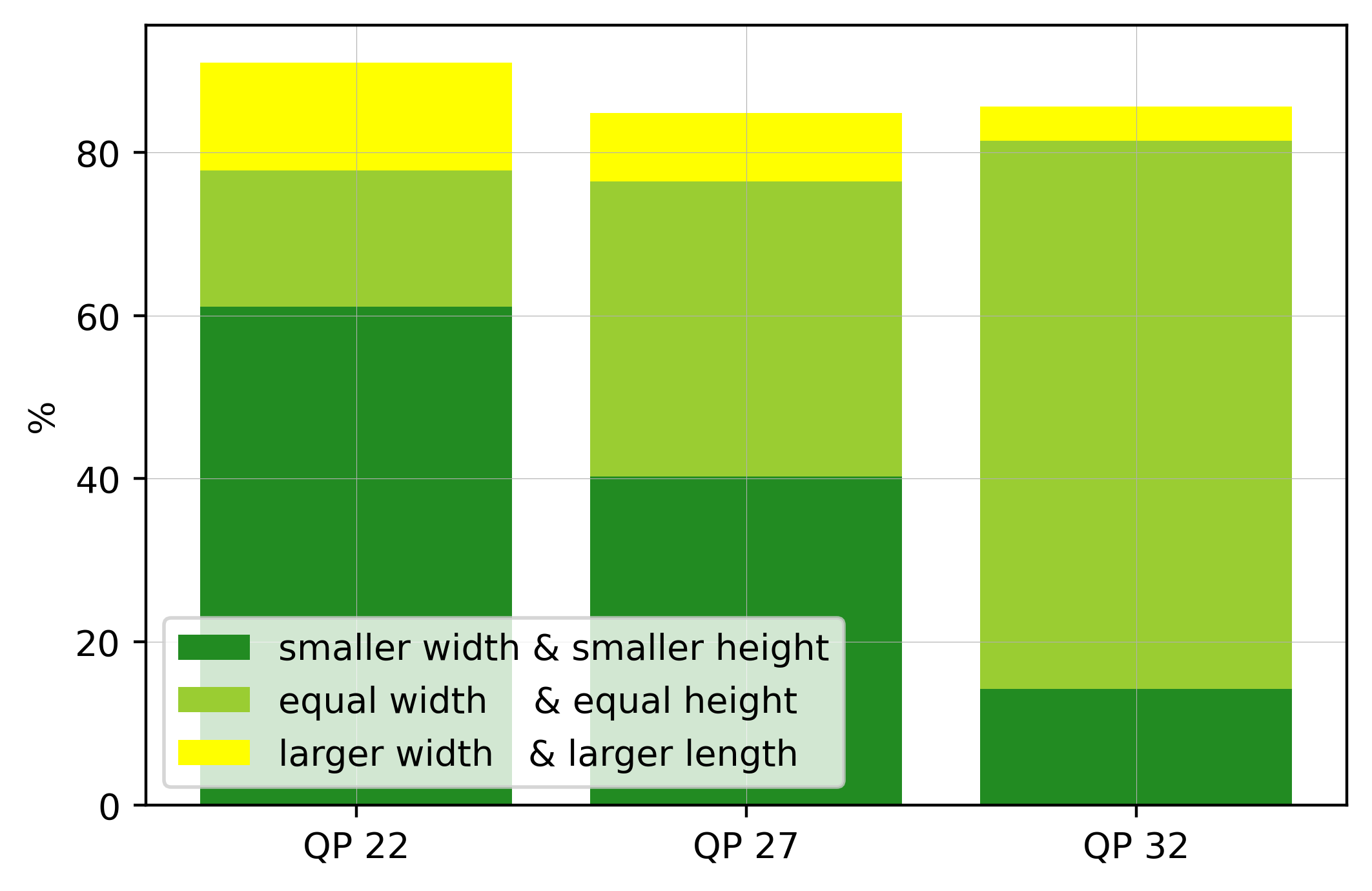}
        \caption{Reference = QP 37}
        \label{fig:similarity_a}
    \end{subfigure}
    \begin{subfigure}{0.24\textwidth}
        \centering
        \includegraphics[width=\textwidth]{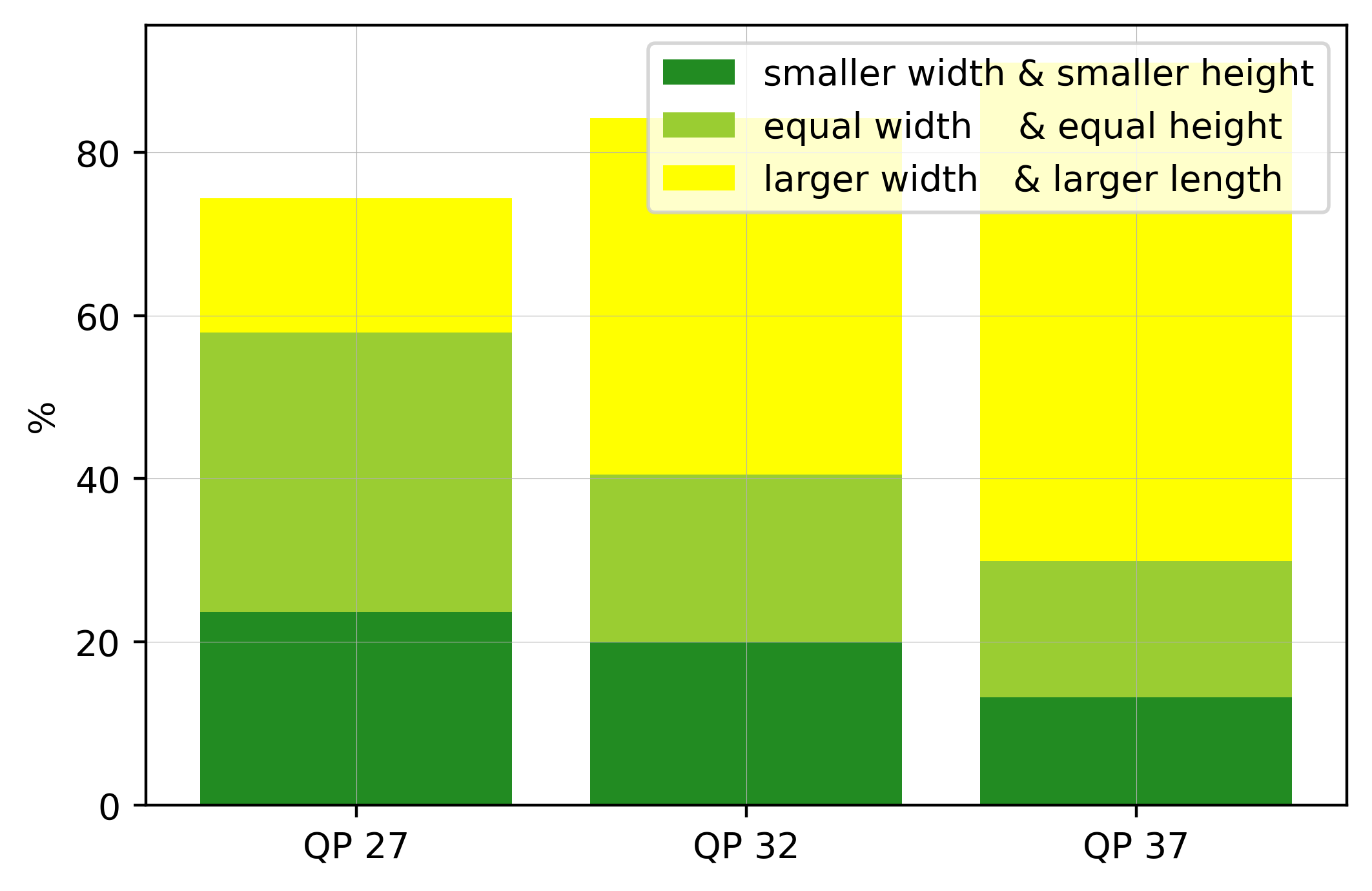}
        \caption{Reference = QP 22}
        \label{fig:similarity_b}
    \end{subfigure}
    \caption{Average percentage of the \gls{cu}s with \gls{cu} height and width larger, equal to, or smaller than the reference encoding.}
    \label{fig:similarity}
    \vspace{-1em}
\end{figure}

Despite the above complications, multi-rate \gls{vvc} encoded representations of a video content can still carry a significant level of redundancies. To demonstrate this, 
we encoded 18 sequences with \gls{ra} coding configurations at \gls{qp}s of 22, 27, 32, 37. These sequences include
six from class A of the VVC CTC~\cite{jvet_ctc} and 12 from the
Inter4K dataset~\cite{stergiou_adapool_2023}. To select 12 sequences out of the 1000 sequences available in the Inter4K dataset, we compared their spatial and temporal complexities using \gls{vca}~\cite{menon_vca_2022}. Based on this comparison, we divided them into four clusters as follows: low spatial and low temporal complexity, low spatial complexity and high temporal complexity, high spatial complexity and low temporal complexity, and high spatial complexity and high temporal complexity. From each cluster, we then randomly selected three sequences for further evaluation (sequences 10, 47, 56, 176, 186, 339, 497, 500, 606, 646, 666, 754). \gls{qp} 22 and 37 were used as reference for comparisons of \gls{cu} sizes in non-reference \gls{qp}s. Fig.~\ref{fig:similarity} (a) and (b) show, respectively, what percentage of encoded co-located \glspl{cu} have \emph{both} \gls{cu} width and height smaller than \gls{qp} 37 or larger than \gls{qp} 22. 

\begin{figure}[!h]
    \centering
    \includegraphics[width=0.7\linewidth]{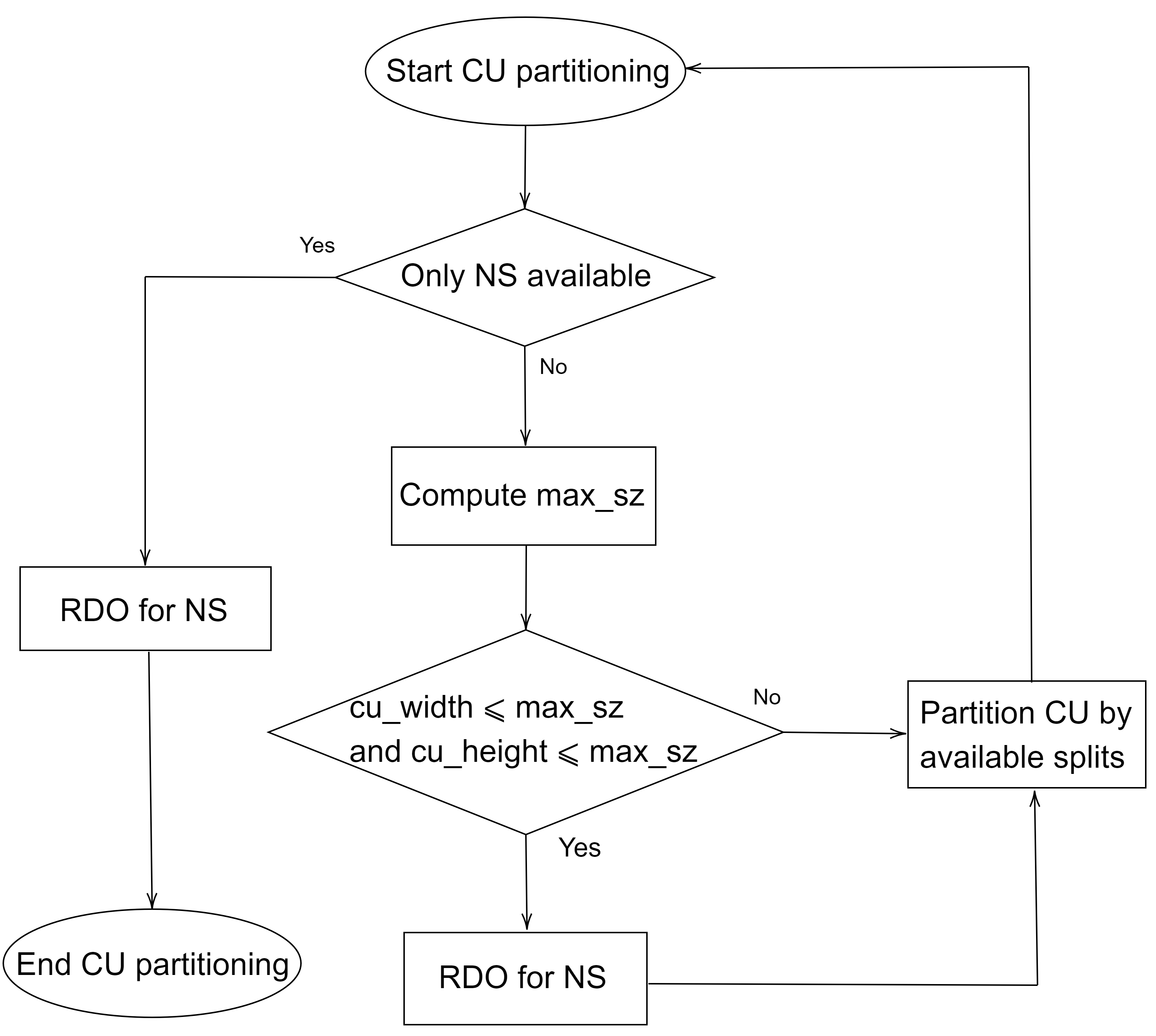}
    \caption{Diagram of \gls{cu} size based fast partitioning.}
    \label{fig:diagram}
    \vspace{-0.5em}    
\end{figure}

When the reference encoding is set to \gls{qp} 37, it is observed that a smaller portion of \gls{cu}s in the dependent encodings have larger width and height sizes compared to their co-located \gls{cu}s in the reference encoding (yellow bars in Fig.~\ref{fig:similarity_a}). The closer the \gls{qp} of the dependent encoding to the reference encoding, the larger the portion of \gls{cu}s that have the same size as the co-located \gls{cu} in the reference encoding. Therefore, when the reference encoding is set to \gls{qp} 37, \gls{rdo} can be skipped for co-located \gls{cu}s that have a width and height larger than the reference \gls{cu}. On the other hand, when the reference encoding is set to \gls{qp} 22,  a smaller portion of \gls{cu}s in the dependent encoding have smaller width and height sizes compared to their co-located \gls{cu}s in the reference encoding (dark green bars in Fig.~\ref{fig:similarity_b}). Therefore, when the reference encoding is set to \gls{qp} 22, \gls{rdo} can be skipped for co-located \gls{cu}s that have a smaller width and height than the reference \gls{cu}. Please note that the cumulative percentages of \gls{cu}s do not reach 100\% because the cases where \gls{cu}s have smaller widths but larger heights or larger widths but smaller heights have not been considered in the evaluations.


\begin{figure}[!t]
    \centering
    \includegraphics[width=0.6\linewidth]{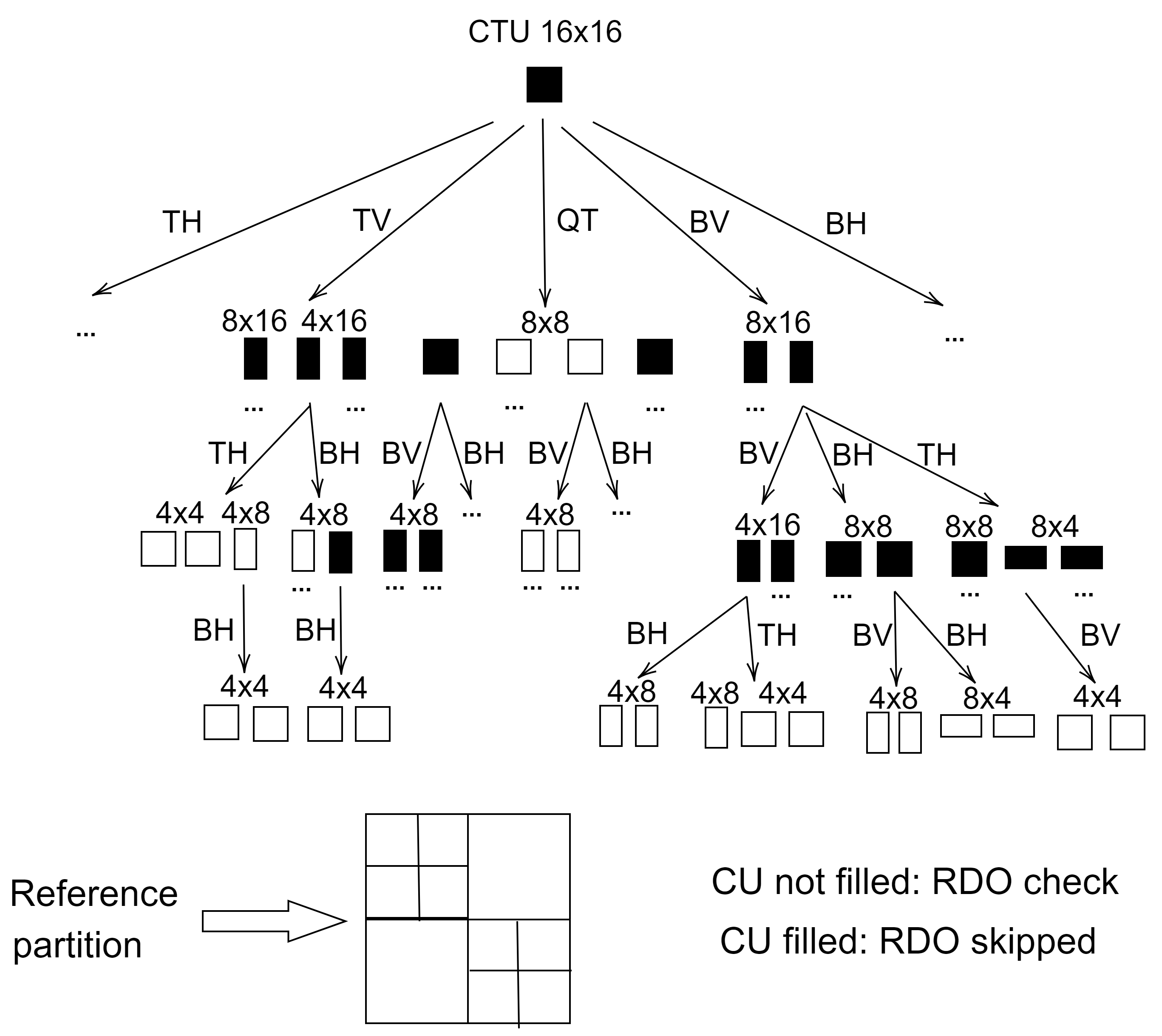}
    \caption{Fast approach involved partition search.}
    \label{fig:partitioning}
    \vspace{-1em}
\end{figure}

\begin{figure*}[!t]
    \centering
    \begin{tabular}{ccc}
            \includegraphics[width=0.25\textwidth]{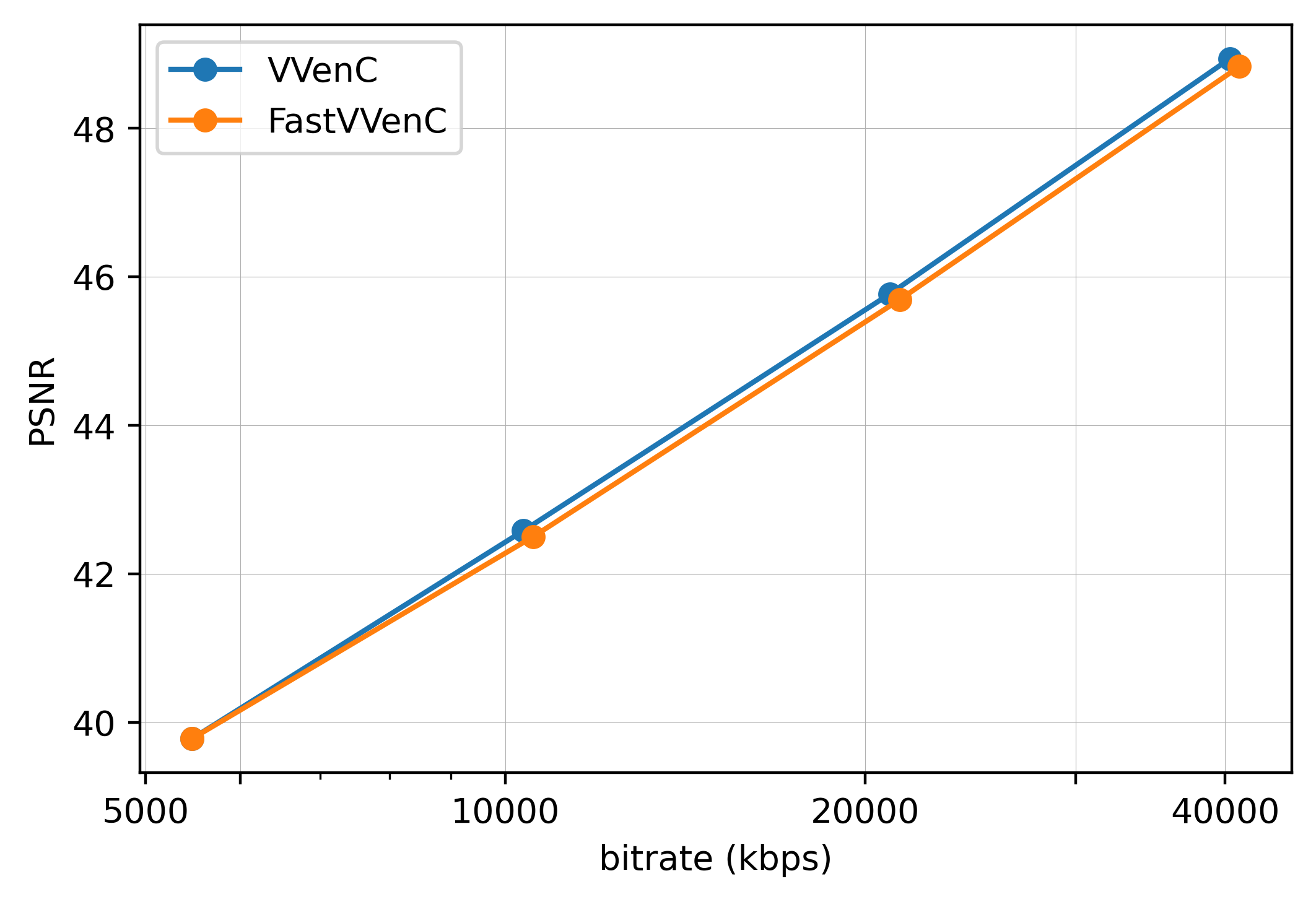}            
         &  
            \includegraphics[width=0.25\textwidth]{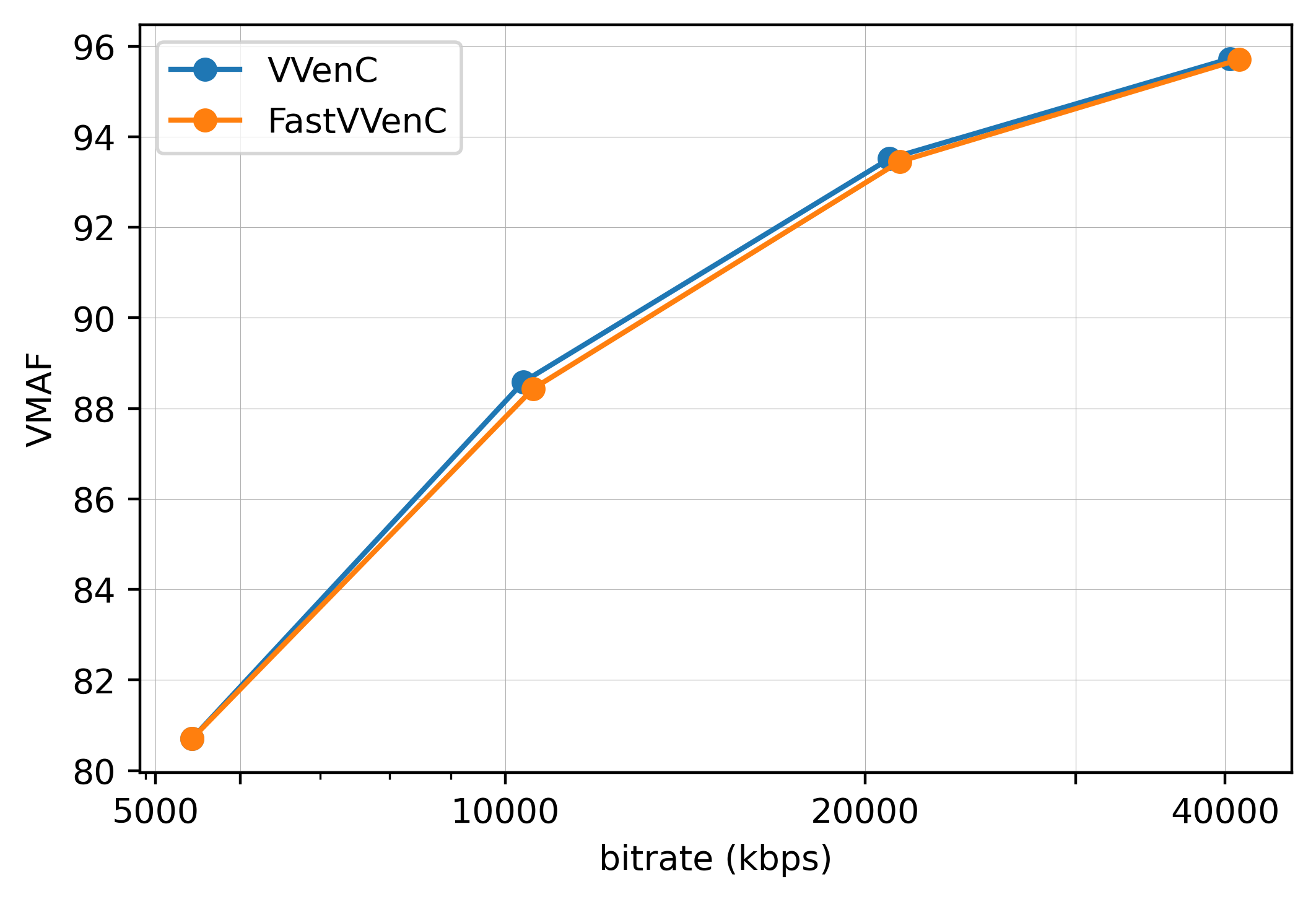}            
        &
            \includegraphics[width=0.25\textwidth]{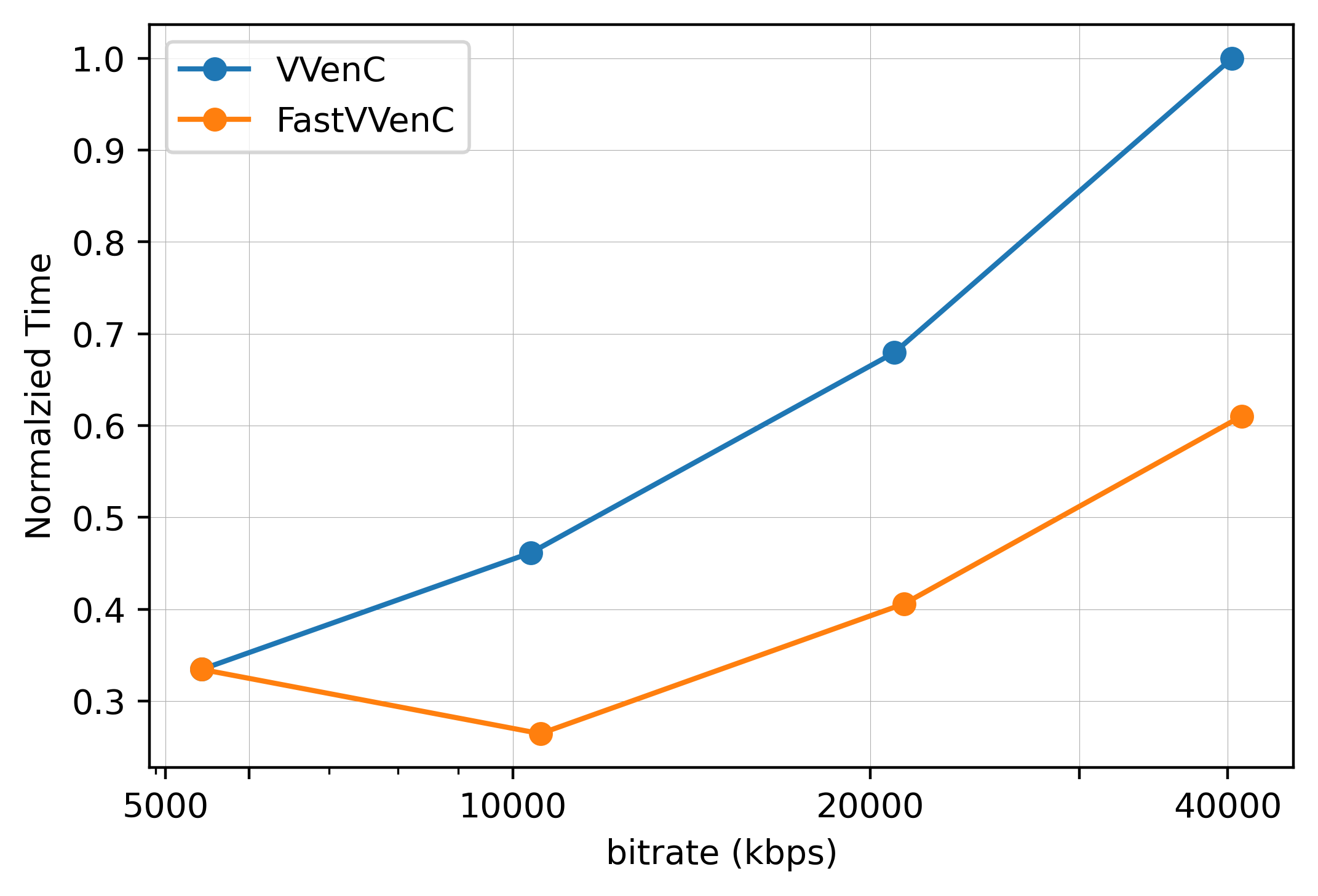}        
        \\
        (a) & (b) & (c)
    \end{tabular}
    \caption{The comparison between the standalone VVenC and the proposed fast multi-rate encoding.}
    \label{fig:results}
    \vspace{-1em}    
\end{figure*}

\section{A fast multi-rate encoding scheme for VVC}
\label{sec:3}
Based on the experiments presented previously, a method is proposed to exploit the redundancy in the \gls{cu} sizes with respect to a reference-coded bitrate. Selecting the appropriate reference encoding for reusing its information to accelerate the encoding of dependent representations is indeed a challenging task and can vary depending on the specific application. In the context of live video streaming, it is commonly observed that the highest quality representation requires the longest encoding time, followed by lower quality representations as the quality decreases. Given the significant impact of encoding time on the latency of live video streaming, it is advantageous to select a representation with a lower encoding time (lower quality) as a reference for accelerating the encoding of representations with higher encoding time (higher quality). By doing so, the maximum and overall encoding process can be expedited, reducing the latency and enhancing the potential for transitioning towards real-time streaming with VVC. Therefore, in this paper, we propose selecting the representation with the lowest encoding time (\gls{qp} 37) as the reference encoding. By reusing its information, we aim to accelerate the encoding process of the dependent representations (\gls{qp} 32, \gls{qp} 27, and \gls{qp} 22).

When a reference \gls{ctu} of 128$\times$128 is encoded at \gls{qp} 37, its size map is calculated to contain the maximum width and height of \gls{cu}s. Since the smallest \gls{cu} size is 4$\times$4, each 4$\times$4 sub-block is assigned the maximum width and height of the \gls{cu} it belongs to. This process results in a map of size 32$\times$32 for each \gls{ctu}. With a reference encoding of \gls{qp} 37 and a reference size map of 32$\times$32, the algorithm for accelerating the encoding of co-located dependent \gls{cu}s is illustrated in Fig.~\ref{fig:diagram}.

Firstly, possible split types of current \gls{cu} are investigated. If no split types other than \gls{ns} are available, the fast partitioning approach will not be applied to current \gls{cu}. The encoding of current \gls{cu} is directly processed and the \gls{cu} partitioning is terminated. If other split types are possible, we obtain \emph{max\_sz} in Fig.~\ref{fig:diagram} by getting the maximum value of elements of the size map inside current \gls{cu}. If the width and height of the current \gls{cu} are both smaller or equal to \emph{max\_sz}, 
we do the \gls{rdo} for current \gls{cu}. Otherwise, the encoding of \gls{cu} is skipped and the \gls{cu} is further split to sub-\gls{cu}s. A part of \gls{rdo}s of \gls{cu} is leaped by applying our fast partitioning method.

To better illustrate the effect of the acceleration achieved by our algorithm, we present the partition search results with our fast approach applied in Fig.~\ref{fig:partitioning}. To simplify the partition tree, we have set the \gls{ctu} size to 16$\times$16 and the \gls{qt} split is disabled for 8$\times$8 \gls{cu}. Meantime, the maximum partitioning depth is 3. For better visualization, repetitive branches are not presented in the figure. \gls{cu}s with \gls{rdo} skipped are filled with black. If we proceed with a larger \gls{ctu} size, the proportion of skipped \gls{cu}s would be larger. 


\section{Experimental results}
\label{sec:4}

To evaluate the proposed algorithm, we performed encodings using the \gls{vvenc} with both the standalone and modified versions of the encoder on the same set of sequences and with the same encoding configuration as mentioned in Section~\ref{sec:3}. For each encoding, we recorded the bitrate, and encoding time, and computed objective metrics, including \gls{psnr} and \gls{vmaf}. Fig.~\ref{fig:results} shows the \gls{psnr} vs. bitrate, \gls{vmaf} vs. bitrate, and normalized time vs. bitrate for the video sequence 754 from Inter4K dataset as an example. It is observed that the rate-distortion performance remains approximately the same while the encoding time is significantly reduced for the dependent representations. From Fig.~\ref{fig:results}-(c), it can be concluded that not only is the overall encoding time of all representations reduced, but also the maximum encoding time is decreased, which is equivalent to the encoding latency. This is highly beneficial in live video streaming as it ensures efficient and timely processing of the video data. In this example, using the proposed method, the maximum encoding time is reduced by 40\%, as can be seen in Fig.~\ref{fig:results}-(c). The average BD-PSNR and BD-VMAF are achieved at -0.21 and -0.43, respectively, indicating that the proposed fast multi-rate encoding method causes a slight decrease in video quality compared to the standalone \gls{vvenc} at equivalent bitrates. At the same time, the average increase in bitrate in equivalent quality is 5.11\% and 4.82\% using \gls{psnr} and \gls{vmaf} as objective metrics, respectively.  However, the reduction in video quality is minimal, and the benefits of significantly reduced encoding time make the proposed method a valuable option for various applications, especially in live video streaming scenarios where real-time performance is crucial. The proposed fast multi-rate encoding results in an average encoding time reduction of 40\% for all dependent representations. Additionally, it achieves an average maximum encoding time reduction of 39\%, indicating that the encoding latency is reduced by 39\% when using this approach.

\section{Conclusions}
\label{sec:5}
This paper proposes a fast multi-rate encoding approach for \gls{vvc} aimed at reducing the encoding time for video streaming applications. The approach involves encoding a reference representation using a standalone \gls{vvc} encoder. The information obtained from the reference representation is then utilized to accelerate the encoding process for the dependent representations. To ensure compatibility with parallel processing, the representation with minimal encoding time is selected as the reference encoding. This reference representation is then utilized to reduce the encoding time of the longer encoding time representations, aiming not only to reduce the overall encoding time but also to minimize the maximum encoding time, which is equivalent to encoding latency. When a \gls{ctu} is encoded in the reference representation, the \gls{rdo} process is skipped in the co-located \gls{ctu}s in dependent representations for \gls{cu}s that have larger width and height compared to the reference \gls{ctu}. The experimental results demonstrate that the proposed method achieves an average 40\% encoding time reduction, while the quality drop, on average, is only 0.43 \gls{vmaf} when using \gls{vvenc}, an optimized \gls{vvc} implementation.

\section{Acknowledgment}
\vspace{-0.3em}
\urlstyle{same}
\small{The financial support of the Austrian Federal Ministry for Digital and Economic Affairs, the National Foundation for Research, Technology and Development, and the Christian Doppler Research Association is gratefully acknowledged. Christian Doppler Laboratory ATHENA: \url{https://athena.itec.aau.at/}.}

\balance
\bibliographystyle{IEEEbib}
\bibliography{refs,ref_hadi}

\end{document}